# Tunable uniform field enhancement in a subwavelength air pillar by photonic doping in epsilon-near-zero medium

Fei Sun[1], Jinyuan Shan[1], and Yichao Liu[1, *]

*1 Key Lab of Advanced Transducers and Intelligent Control System, Ministry of Education and Shanxi Province, College of Electronic Information and Optical Engineering, Taiyuan University of Technology, Taiyuan, 030024 China*

*\* liuyichao@tyut.edu.cn*

**Abstract**

In this study, a novel electric field compressor is proposed by doping a metal-air-metal pillar in epsilon-near-zero medium within a metallic waveguide, which effectively enhances the background electric fields in a sub-wavelength air pillar with high uniformity. The field enhancement factor can be analytically determined through theoretical derivations from Maxwell's equations, reaching its maximum value by appropriately selecting the size of the air pillar. Furthermore, the proposed compressor can achieve a tunable electric field enhancement effect within the deep sub-wavelength air pillar by adjusting the height of the air pillar through movement of two metal pillars inserted into the waveguide. Both theoretical analysis and numerical simulations are employed to validate the performance of the electric field compressor, which exhibits a wide range of tunable field enhancement effect (i.e., continuously adjustable enhancement factor between 20 and 800) with uniformity below 1.5 within the deep sub-wavelength air pillar (wherein the air volume is smaller than $\lambda_0/10 \times \lambda_0/10 \times \lambda_0/16$). Finally, a practical model is proposed for the realization of the electric field compressor in the microwave range, where specially placed metal pillars and wires are incorporated into a rectangular metallic waveguide structure. The field enhancement and tunable effects of this practical model have been verified through simulations.

**Introduction**

Confining electric fields within subwavelength scale has significant applications in various fields, such as improving the quality of biomedical imaging and biosensors [1,2], detecting extremely weak electromagnetic signals (e.g., cosmic background electromagnetic radiation), amplification of field-matter interaction on micro/nano scales [3], surface enhanced Raman scattering [4,5], optical tweezing [6], and quantum computing [7]. Current methods of enhancing electric fields in subwavelength scale primarily involve plasmonic structures with geometrical singularities [8-11], concentrators designed by transformation optics [12-15], and near-zero-refractive-index metamaterials (ZIMs) [16-20]. However, these existing methods cannot obtain a uniform magnification of electric fields within a subwavelength air region. Plasmonic structures can induce a field enhancement effect near the plasmonic tips/gaps, thus failing to provide uniform enhancement factors (i.e., maximum enhanced field occur at geometrical singularities) [8-11]. Electromagnetic concentrators can provide uniform enhanced electric fields within a non-air region, filled with anisotropic high refractive index medium, typically much larger than the wavelength [12-15]. The ZIMs can achieve enhanced electric fields within the ZIMs (not in air) with a non-uniform enhancement factor [16-18]. Therefore, it is still challenging to achieve electric field enhancement in a sub-wavelength air region with high uniformity.

In our previous work, near-zero-index-featured medium is proposed to realize ZIMs equivalently by nonmagnetic normal dielectrics [19,21]. Subsequently, an optical funnel is designed based on this near-zero-index-featured medium to uniformly enhance the electric field in the air region [20]. However, the optical funnel requires precisely gradient refractive index media with high refractive indices, which is difficult to be fabricated experimentally. Consequently, effective methods for achieving high uniform electric field enhancement in a subwavelength air region by homogeneous media need to be further explored. In recent years, photonic doping of ZIM has been used to realize various novel electromagnetic phenomenon [22], such as perfect coherent absorption [23], dispersion coding [24], and unusual percolation [25]. In this study, a novel electric field compressor (EFC) is designed by doping a metal-air-metal pillar in epsilon-near-zero (ENZ) medium, which can create an enhanced electric field within a subwavelength air pillar with high enhancement factor and uniformity.

**Results and discussions**

The basic structure of the proposed EFC is shown in Fig. 1. The whole structure is a rectangular waveguide with cross-sectional size of $a \times H$ (*y-z* plane). The top/bottom sides are designed as perfect electric conductors (PEC), the front/back/right sides are designed as perfect magnetic conductors (PMC). The left side remains open, serving as both an input and output port, as the right side is blocked by PMC. Region 1 of the waveguide is filled with air, and region 2 of the waveguide is filled with ENZ medium ($\varepsilon_{ENZ} \approx 0$, $\mu = 1$) doped by a rectangular metal-air-metal pillar. The rectangular air impurity pillar between the two metal pillars has a cross-sectional size of $s \times s$ (*x-y* plane), and height of $h$ (*z*-direction). For the convenience of the following analysis, the relative permittivity of the rectangular air impurity pillar is denoted by $\varepsilon_d$.

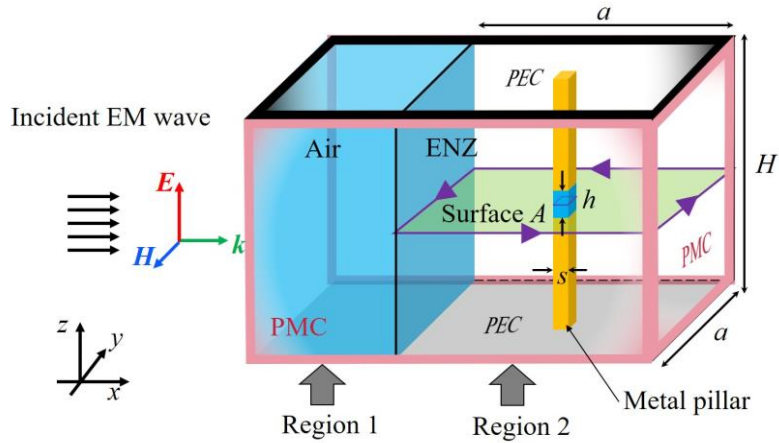

Fig. 1. Basic structure of the EFC, which is a waveguide structure composed of two regions. The region 1 (colored blue) and region 2 (transparent) are air and ENZ medium, respectively. The two yellow rectangular pillars are the metal impurities, and the blue rectangular pillar is the air impurity.

The pink framed surfaces are PMC and the black framed surfaces are PEC.

Assuming a time-harmonic transverse electromagnetic (TEM) wave with electric field $\boldsymbol{E}_i = \hat{z}E_0 e^{ik_0 x}$ incidents from the left port of the waveguide (see Fig. 1). Here, $E_0$ and $k_0$ are amplitude and wavenumber of the incident electromagnetic wave, respectively. The total fields in region 1 can be expressed as,

$$\begin{cases} \mathbf{E}^{(1)} = \mathbf{E}_i + \mathbf{E}_r = \hat{z}E_0 \left( e^{ik_0 x} + Re^{-ik_0 x} \right) \\ \mathbf{H}^{(1)} = -\frac{k_0}{\omega\mu_0}\mathbf{E}_i + \frac{k_0}{\omega\mu_0}\mathbf{E}_r = \hat{y}\frac{k_0 E_0}{\omega\mu_0}\left( -e^{ik_0 x} + Re^{-ik_0 x} \right) \end{cases}, \quad (1)$$

where $R$ is the reflection coefficient, $\hat{y}$ and $\hat{z}$ denote unit vectors in $y$ and $z$ direction, respectively. The $x = 0$ plane is chosen as the Air/ENZ boundary, and the time harmonic factor $e^{-i\omega t}$ is omitted for convenience. According to Maxwell's equations, the loop integration of magnetic field along the boundary of surface $A$ (denoted by the green surface) is related to the surface integration of electric fields,

$$\int_{\partial A} d\mathbf{l} \cdot \mathbf{H} = -i\omega\varepsilon_0\varepsilon_d \int_{A_d} d\mathbf{S} \cdot \mathbf{E} - i\omega\varepsilon_0\varepsilon_{ENZ} \int_{A_{ENZ}} d\mathbf{S} \cdot \mathbf{E}, \quad (2)$$

where $A_{ENZ}$ and $A_d$ ($A_{ENZ} + A_d = A$) denote the cross section of the ENZ medium and air impurity pillar, respectively. As the tangential components of magnetic fields vanish at the PMC boundary, the left-hand side of Eq. (2) is equal to $aE_0 k_0(1-R)/\omega\mu_0$. Due to the negligible contribution of the second term on the right side of Eq. (2) when $\varepsilon_{ENZ} \to 0$, electric flux density mainly concentrates inside the air impurity. This is the physical mechanism by which the designed EFC has the ability of squeezing electric fields within the rectangular air impurity pillar of the ENZ medium. In this case, the average electric fields $E_{avg}$ perpendicular to the air surface $A_d$ can be calculated as,

$$E_{avg} = i\frac{a(1-R)}{\varepsilon_d k_0 s^2} E_0. \quad (3)$$

To describe the degree of the enhanced electric field in the air pillar compared to the incident electric field, we define the average enhancement factor of electric field as

$$M_{avg} = \left|\frac{E_{avg}}{E_0}\right| = \frac{\lambda_0 a |1-R|}{2\pi\varepsilon_d s^2}, \quad (4)$$

where $\lambda_0 = 2\pi/k_0$ is the wavelength of the incident wave. In this study, the central working wavelength is designed as $\lambda_0 = 0.3$ m. Given the incident wavelength $\lambda_0$, then the average enhancement factor is determined by the length of the ENZ medium in $y$-direction (i.e., $a$), the cross-sectional size of the air impurity pillar (i.e., $s^2$), and the reflection coefficient $R$. Note that the reflection coefficient is highly dependent on the air pillar. Although the amplitude of reflection coefficient is equal to one due to the total reflection of electromagnetic waves by the right PMC boundary, the phase of reflection coefficient varies as the sizes or permittivity of the impurity change. To achieve the maximum enhancement factor, the reflection phase should be 180 degree according to Eq. (4), i.e., $R = -1$.

To understand how the impurity influence the reflection phase, we firstly assume the case without impurity, i.e., the air impurity is replaced by background ENZ medium (i.e., $\varepsilon_d \rightarrow 0$). Eq. (2) indicates that the magnetic fields become zero at the $x = 0$ plane, thus leading to zero reflection phase. Then, keeping other parameters unchanged (i.e., $a = 10\lambda_0/3$, $H=0.05a$, $s = 0.02a$, $h = 0.2H$ and $\lambda_0 = 0.3$m), we gradually increase the permittivity of the impurity. In this case, the reflection phase decreases accordingly and 180-degree reflection phase can be achieved at a proper permittivity, specifically corresponding to the resonance peak position of the red dotted line in Fig. 2(a). The simulated enhancement factor and calculated value using Eq. (4) are plotted by the blue and black curves in Fig. 2(a), respectively. Please note we use $\varepsilon_{ENZ} = 0.001$ instead of exactly zero in all simulations, which leads to the non-zero integration contribution of the second term on the right side of Eq. (2), thus the calculated value is slightly larger than the simulated one. This integration contribution can be neglected when the permittivity of the impurity is much larger than the permittivity of the ENZ medium (e.g., for $\varepsilon_d > 1$, the black and blue curves essentially overlap in Fig. 2(a)). However, when the permittivity of the impurity is comparable to that of the ENZ medium, the second term of the right side of Eq. (2) cannot be neglected as it is comparable to the first term. In this case, the simulated enhancement factor deviates significantly from the calculated value, e.g., the black and blue curves deviate significantly for $\varepsilon_d < 0.5$ in Fig. 2(a).

In the subsequent practical application examples, air is selected as the impurity, while keeping both the size of the background ENZ medium and the cross-sectional dimension of the doped metal-air-metal pillar constant. However, the height of the air pillar can be tuned by varying the lengths to which two metal pillars are inserted into the waveguide. Therefore, we investigate how the height of the air pillar affects the reflection phase and its role in tuning the average enhancement factor. To find the maximum enhancement factor within air pillar, we change the height of air pillar $h$ from $0.1H$ to $0.3H$, while maintaining other parameters constant (specifically, $a = 10\lambda_0/3$, $H=0.05a$, $s = 0.02a$, and $\lambda_0 = 0.3$m). In this case, the reflection phase covers a range of 360 degrees, as illustrated by the red dotted line in Fig. 2(b). The calculated enhancement factor using Eq. (4) fits well with the simulated one, which are represented by the black and blue curves, respectively, in Fig. 2(b). At the 180-degree reflection phase, a maximum of the simulated/calculated enhancement factor is observed, where $h=0.177H$ corresponding to the blue/black peak in Fig. 2(b). As the height of metal pillars inserted into the waveguide can be freely adjusted (i.e., change freely in the $z$-direction), the height of the air pillar $h$ can be tuned without disturbing the ENZ medium, thus achieving a tunable enhancement factor. Moreover, the calculated and simulated results in Fig. 2(b) shows that when the length of the air pillar remains constant on the subwavelength order (e.g., $s =\lambda_0/15$), and the height of the air pillar varies within the range of $\lambda_0/60$ to $\lambda_0/20$, the entire spatial volume of the air pillar can be on the deep subwavelength scale (minimally achievable at $\lambda_0^3/13500$) with the enhancement factor ranging from 20 to 200. Therefore, the EFC allows for a wide range of tunable enhancement factor in subwavelength-scale air pillar by adjusting the height of the metal pillars inserted into the waveguide.

Although we can tune the height of air pillar to achieve the 180-degree reflection phase, the enhancement factor is still bounded by $\lambda_0 a/\pi s^2$. Therefore, a more efficient way to substantially increase the enhancement factor is to decrease the cross-sectional size of the air pillar, as predicted by Eq. (4). The simulated and theoretical predicted enhancement factors are depicted as blue and black curves in Fig. 2(c), respectively, for varying cross-sectional side length of the air pillar from $0.01a$ to $0.03a$. In each case, the height of air pillar is carefully chosen to make sure the reflection phase to be 180 degree, as illustrated by the pink curve in Fig. 2(c). The difference between the simulated and calculated enhancement factor becomes larger as the cross-sectional side length decreases. The reason is that, as the electric fields are substantially enhanced, the large electric fields are also partially distributed in the ENZ medium near the air pillar, which make a larger contribution to the integration of the right-hand side of Eq. (2) but is neglected in Eq. (4). Both the simulated and calculated results indicate that it is possible to increase the enhancement factor (i.e., achieving the larger electric field in air pillar) by decreasing the cross-sectional size of the air pillar, while simultaneously decreasing the height of the air pillar (to maintain a reflection phase of 180 degree).

Next, we investigate the bandwidth of the EFC, specifically the relationship between the enhancement factor and the incident wavelength $\lambda$. Firstly, we fix the height of the air pillar, and continuously change the incident wavelength from $0.5\lambda_0$ to $1.5\lambda_0$. The simulated enhancement factor is shown as the green curve in Fig. 2(d), revealing a relatively broad bandwidth for the field enhancement effect, approximately 10.5% calculated by $\delta(1/\lambda)/(1/\lambda_0)$. As the height of the air pillar can be tuned, the field enhancement effect can be achieved in a broad band. By tuning the height of air pillar, as shown in the pink curve in Fig. 2(d), the enhancement factor (shown as blue curve in Fig. 2(d)) can exceed 100 if the incident wavelength is larger than $0.7\lambda_0$, which means a broadband enhancement can be achieved. With proper chosen height of the air pillar, the enhancement factor increases approximately linearly with the incident wavelength, which coincide with theoretical prediction, as depicted by the black curve in Fig. 2(d).

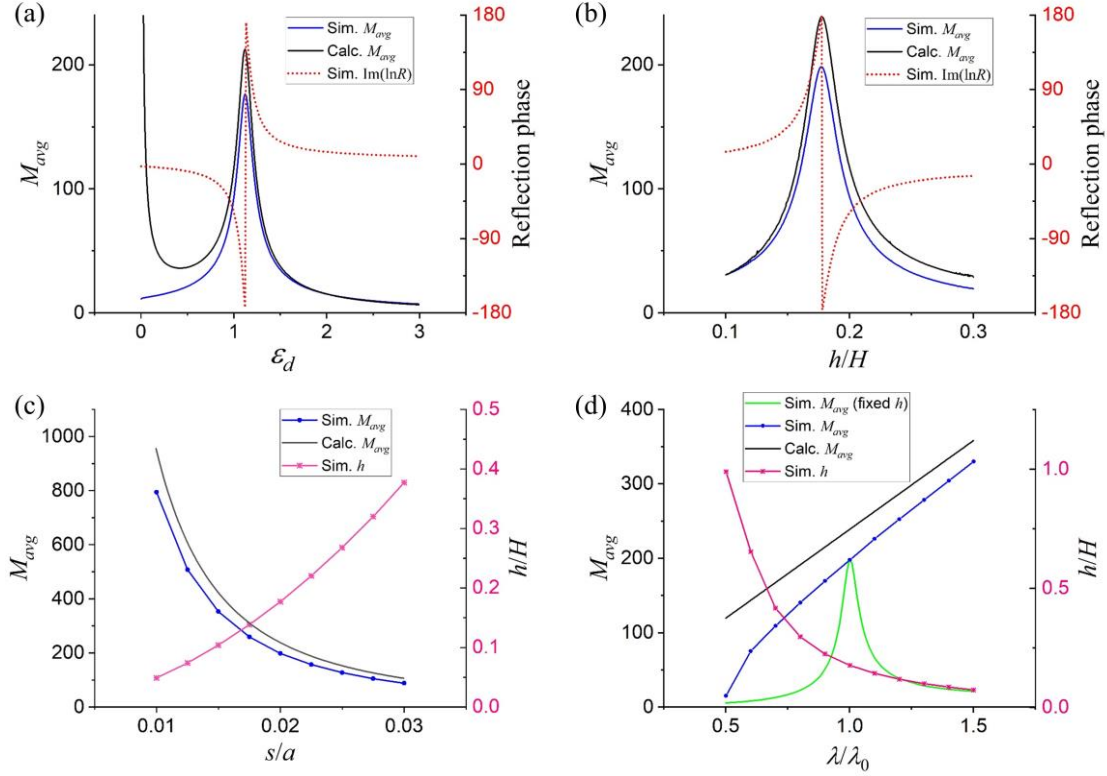

Fig. 2. (a) Enhancement factor (blue and black curves on the left axis) and reflection phase (red dots on the right axis) for varied permittivity of the impurity. (b) Enhancement factor (blue and black curves on the left axis) and reflection phase (red dots on the right axis) for varied impurity height. (c) Enhancement factor (blue and black curves on the left axis) for varied cross-sectional side length of the air pillar with optimized height (pink curve on the right axis). (d) The relation between enhancement factor and varied incident wavelength for the fixed height of the air pillar (green curve) and optimized height of the air pillar (blue and black curves). The corresponding optimized height of the air pillar is depicted by pink curve. The geometrical parameters of the EFC are set as $a = 10\lambda_0/3$, and $H=0.05a$ for (a)-(d); $s = 0.02a$ for (a), (b) and (d); $h = 0.2H$ for (a) and $h = 0.177H$ for the green curve in (d). The central working wavelength is $\lambda_0=0.3$ m

To check the uniformity of the enhanced electric fields inside the air pillar, we choose the cross-sectional side length $s$ and height $h$ of the air pillar to be the same as the those used in Fig. 2(c). We define surface uniformity $\sigma_S$ and volume uniformity $\sigma_V$ as the standard deviation of the electric fields on the central surface (x-y plane) and the whole volume of the air pillar [20], respectively, i.e., $\sigma_S = \langle E^2 \rangle_S - \langle E \rangle_S^2$, $\sigma_V = \langle E^2 \rangle_V - \langle E \rangle_V^2$, where $\langle \rangle_S$ and $\langle \rangle_V$ mean averaging over the surface or the volume, respectively. Simulation results in Fig. 3(a) show both the surface uniformity and volume uniformity are below 1.5 for all the enhancement factors (from 80 to 800), which verify the good uniformity of the electric fields inside the air pillar. As shown in Fig. 3(a), when the sizes of the air pillar are chosen as $0.01a$ (length) × $0.01a$ (width) × $0.0024a$ (height), a uniform 800-fold enhancement of the background electric field with a uniformity less than 1.5 can be achieved within a deep subwavelength three-dimensional air pillar (i.e., $\lambda_0/30 \times \lambda_0/30 \times \lambda_0/123$). If the sizes of the air pillar increase to $0.03a$ (length) × $0.03a$

(width) × 0.0189$a$ (height), the uniformity will be improved (i.e., below 1.2) and the enhancement factor decreases to 90. In this case, the field patters in three cross-sectional views are shown in Fig. 3(b)-3(d), where the uniform enhancement effect inside the air pillar can still be clearly observed. In the following implementation plan for the proposed EFC, considering some actual limitations (such as machining accuracy), we choose an EFC with air pillar $s=0.02a$ in Fig. 3(a) (still corresponding to the deep subwavelength volume, i.e., $\lambda_0/15 \times \lambda_0/15 \times \lambda_0/34$), which can achieve an enhancement factor of 200 and uniformity below 1.3 for the subsequent practical model design.

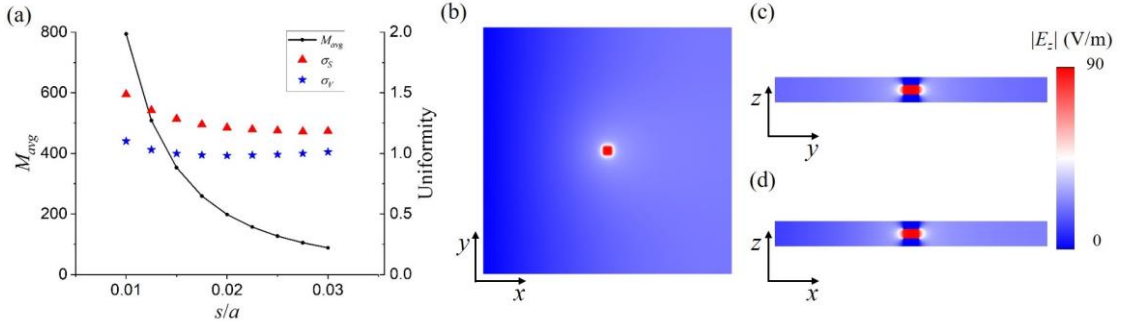

Fig. 3. (a) Enhancement factor and surface/volume uniformity with varied cross-sectional side length of the air pillar. (b)-(d) The simulated amplitude of the electric field's $z$ component within the region 2 of the EFC when a TEM wave with unit amplitude incident onto the EFC at the working wavelength. (b)-(d) represent cross-sectional slices along the $z$, $x$, and $y$ axes, respectively, where the side length of air pillar is chosen as $s = 0.03a$.

To realize the proposed EFC, the key elements to implement are PEC, PMC and ENZ medium. Here, we take the example of microwaves with a central working wavelength of $\lambda_0=3$m and provide a specific scheme for implementing the EFC. In the microwave range, most metals in nature perform as PEC, which we choose here as copper [26,27]. For PMC, there are many equivalent implementation methods [28-30]. Here, we use PEC and a protruding air plate with a quarter wavelength to mimic PMC (see Fig. 4(a) and 4(b)) [31]. For ENZ medium, we use metal wire array to achieve near zero permittivity. As shown in the top view in Fig. 4(a), each metal wire with a radius of $r$ (colored red) is located within a square air unit with a period of $p$, forming a periodic structure to achieve an ENZ medium. Note that there are no metal wires in the position of the metal pillars (colored yellow). The permittivity of metal wire array can be described by Drude model [25],

$$\varepsilon_{\mathit{eff}} = 1 - \frac{\omega_p^2}{\omega(\omega+i\gamma)}, \qquad (5)$$

where $\omega_p = \frac{c}{p}\sqrt{\frac{2\pi}{\ln(p/r)}}$ is the plasma frequency, and $\gamma = \frac{\varepsilon_0 p^2 \omega_p^2}{\pi r^2 \sigma}$ is the damping coefficient. By properly chosen the radius of the metal wire $r$ and the period of the wire array $p$, $\varepsilon_{ENZ} \to 0$ can be achieved at the designed working wavelength $\lambda_0$, which is shown in Fig. 4(c).

As shown in the side view in Fig. 4(b), metal pillars performing as impurities (colored yellow) are inserted into the waveguide by drilling holes from the top and bottom of the waveguide. By changing the length of the copper pillars inserted into the waveguide, the height of the air pillar $h$ can be directly tuned. Simulations are performed to verify the tunability of the enhancement factor after the ideal EFC model is replaced by practical EFC model in Fig. 4(a). As shown in Fig. 4(d), when the wavelength is fixed to $\lambda_0$, both enhancement factors for the ideal EFC model using PMC (colored black) and the practical EFC model using PEC/air (colored blue) can be tuned by varying the height of air pillar from $0.1H$ to $0.3H$. A similar tunable enhancement effect is observed in both ideal and practical EFC models.

Then, we study the bandwidth of the practical EFC model. If the size of the air pillar and other parameters of the practical EFC model are fixed, when the incident wavelength varies from $0.5\lambda_0$ to $1.5\lambda_0$ the enhancement factor reach its maximum at the designed central working wavelength $\lambda_0$. However, at other frequencies, the enhancement factor exhibits significant fluctuations, as illustrated by the red curve in Fig. 4(e). Therefore, the bandwidth of the practical EFC model is narrow, as shown in the inset of Fig. 4(e), which is mainly caused by the material dispersion of the metal wire.

From the above analysis, we find the tunable field enhancement effect of the practical EFC model can still exist but with a narrowed working bandwidth. In the practical EFC model, the electric field pattern at the working wavelength and with proper height of air pillar is shown in Fig. 4(f), which shows similar field enhancing effect as the ideal EFC in Fig. 3(b).

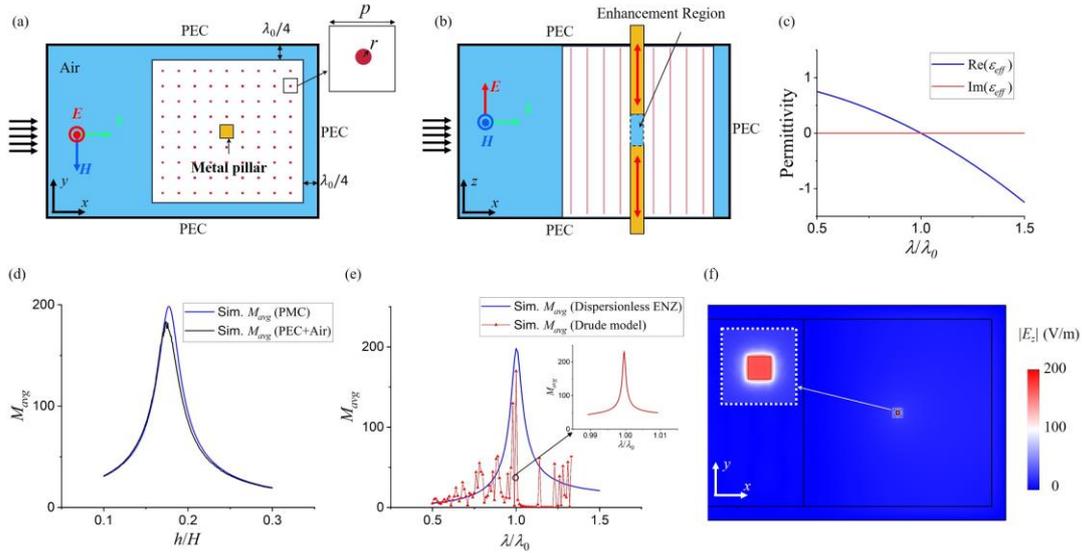

Fig. 4. A practicable structure to realize the EFC, (a) cross-sectional view in $x$-$y$ plane, where the cross-sectional size of the metal pillar is chosen as $s = 0.02a$; (b) cross-sectional view in $x$-$z$ plane, where the metal pillars can move in $z$ direction freely. (c) Permittivity dispersion of the metal wire array calculated by Drude model, where $r = 85.8\times10^{-6}\ a$, $p = a/21$, $\sigma = 5.96\times10^7\ \Omega^{-1}\cdot m^{-1}$, and the working wavelength is chosen as $\lambda_0 = 0.3$m. (d) Enhancement factor with varied height of the air pillar for ideal EFC model (blue) and practical EFC model (black). (e) Enhancement factor with varied incident wavelength for ideal EFC model (blue) and practical EFC model (red). (f) Electric field pattern at working wavelength $\lambda_0$ for the practical EFC model.

**Conclusion**

In conclusion, an EFC is proposed by doping a metal-air-metal pillar in ENZ medium, which can provide uniform enhancement effect of background electric fields inside a subwavelength air pillar between the two metal pillars. The EFC can be realized by placing metal wire arrays (ENZ medium) in a rectangular metallic waveguide and drilling holes at symmetric positions above and below the waveguide to insert two metal pillars. The physical mechanism of the EFC's uniform field enhancement effect originates from the field squeezing property of the surrounded ENZ medium that encapsulates the air pillar. Through some theoretical derivations from Maxwell's equations, we find that the electric field enhancement factor of the proposed EFC can be analytically determined, which is dependent on the central operating wavelength, dimensions of the air pillar, and the sizes of the waveguide. Both theoretical analysis and numerical simulations show that the maximum enhancement factor of the EFC is inversely proportional to the cross-sectional size of the air pillar. Additionally, the relationship between the enhancement factor and the height of the air pillar resembles a resonance curve, exhibiting an optimal height value to achieve the highest field enhancement factor. Moreover, by appropriately changing the size of the air pillar (e.g., adjusting the metal pillars inserted into the waveguide), the ideal EFC can achieve a tunable electric field enhancement effect within the deep sub-wavelength air pillar, exhibiting a wide range of enhancement factors (e.g., $M_{\text{avg}}$ is continuously adjustable between 20 and 800) and a high level of uniformity (i.e., uniformity is below 1.5). Finally, we present the practical implementation model of the EFC in the microwave range (i.e., $\lambda_0$=3m) and verify its tunable field enhancement effect through numerical simulations. The proposed EFC has a wide range of applications in the fields such as detecting extremely weak electromagnetic signal, quantum computing, sub-wavelength electromagnetic/biological manipulation and sensing.

**References**

[1] A. V. Kabashin, P. Evans, S. Pastkovsky, W. Hendren, G. A. Wurtz, R. Atkinson, R. Pollard, V. A. Podolskiy, and A. V. Zayats, Plasmonic nanorod metamaterials for biosensing, Nat. Mater. **8**, 867 (2009).

[2] A. G. Brolo, Plasmonics for future biosensors, Nat. Photonics **6**, 709 (2012).

[3] X. D. Chen, E. H. Wang, L. K. Shan, C. Feng, Y. Zheng, Y. Dong, G. C. Guo, and F. W. Sun, Focusing the electromagnetic field to 10− 6 λ for ultra-high enhancement of field-matter interaction, Nat. Commun. **12**, 6389 (2021).

[4] F. Liu, B. Song, G. Su, O. Liang, P. Zhan, H. Wang, W. Wu, Y. Xie, and Z. Wang, Sculpting extreme electromagnetic field enhancement in free space for molecule sensing, Small **14**, 1801146 (2018).

[5] Y. Hang, J. Boryczka, and N. Wu, Visible-light and near-infrared fluorescence and surface-enhanced Raman scattering point-of-care sensing and bio-imaging: A review,

Chem. Soc. Rev. **51**, 329 (2022).
[6] M. L. Juan, M. Righini, and R. Quidant, Plasmon nano-optical tweezers, Nat. Photonics **5**, 349 (2011).
[7] J. C. Bardin, D. H. Slichter, and D. J. Reilly, Microwaves in quantum computing, IEEE Microw. Mag. **1**, 403 (2021).
[8] J. Hong, S. J. Kim, I. Kim, H. Yun, S. E. Mun, J. Rho, and B. Lee, Plasmonic metasurface cavity for simultaneous enhancement of optical electric and magnetic fields in deep subwavelength volume, Opt. Express **26**, 13340 (2018).
[9] J. A. Schuller, E. S. Barnard, W. Cai, Y. C. Jun, J. S. White, and M. L. Brongersma, Plasmonics for extreme light concentration and manipulation, Nat. Mater. **9**, 193 (2010).
[10] M. K. Kim, H. Sim, S. J. Yoon, S. H. Gong, C. W. Ahn, Y. H. Cho, and Y. H. Lee, Squeezing photons into a point-like space, Nano Lett. **15**, 4102 (2015).
[11] Y. Lu, G. L. Liu, J. Kim, Y. X. Mejia, and L. P. Lee, Nanophotonic crescent moon structures with sharp edge for ultrasensitive biomolecular detection by local electromagnetic field enhancement effect, Nano Lett. **5**, 119 (2005).
[12] M. Y. Zhou, L. Xu, L. C. Zhang, J. Wu, Y. B. Li, and H. Y. Chen, Perfect invisibility concentrator with simplified material parameters, Front. Phys. **13**, 1 (2018).
[13] M. Sadeghi, S. Li, L. Xu, B. Hou, and H. Chen, Transformation optics with Fabry-Pérot resonances, Sci. Rep. **5**, 1 (2015).
[14] C. Lan, B. Li, and J. Zhou, Simultaneously concentrated electric and thermal fields using fan-shaped structure, Opt. Express **23**, 24475 (2015).
[15] A. Abdolali, H. Barati Sedeh, and M. H. Fakheri, Geometry free materials enabled by transformation optics for enhancing the intensity of electromagnetic waves in an arbitrary domain, J. Appl. Phys. **127** (2020).
[16] Y. Jin, P. Zhang, and S. He, Squeezing electromagnetic energy with a dielectric split ring inside a permeability-near-zero metamaterial, Phys. Rev. B **81**, 085117 (2010).
[17] M. G. Silveirinha and N. Engheta, Theory of supercoupling, squeezing wave energy, and field confinement in narrow channels and tight bends using ε near-zero metamaterials, Phys. Rev. B **76**, 245109 (2007).
[18] N. M. Litchinitser, A. I. Maimistov, I. R. Gabitov, R. Z. Sagdeev, and V. M. Shalaev, Metamaterials: electromagnetic enhancement at zero-index transition, Opt. Lett. **33**, 2350 (2008).
[19] Y. C. Liu, F. Sun, Y. B. Yang, Z. H. Chen, J. Z. Zhang, S. L. He, and Y. G. Ma, Broadband Electromagnetic Wave Tunneling with Transmuted Material Singularity, Phys. Rev. Lett. **125**, 207401 (2020).
[20] F. Sun, Y. Liu, and Y. Yang, Optical funnel: broadband and uniform compression of electromagnetic fields to an air neck, Photonics Res. **9**, 1675 (2021).
[21] Y. Liu, F. Sun, Y. Ma, Z. Wang, and Y. Liu, Near-zero-index-featured multi-band highly directional radiator with large Purcell factors, Results Phys. **40**, 105875 (2022).
[22] I. Liberal, A. M. Mahmoud, Y. Li, B. Edwards, and N. Engheta, Photonic doping of epsilon-near-zero media, Science **355**, 1058 (2017).
[23] J. Luo, B. Liu, Z. H. Hang, and Y. Lai, Coherent perfect absorption via photonic doping of zero‐index media, Laser Photon. Rev. **12**, 1800001 (2018).
[24] Z. Zhou, H. Li, W. Sun, Y. He, I. Liberal, N. Engheta, Z. Feng, and Y. Li, Dispersion

coding of ENZ media via multiple photonic dopants, Light: Sci. Appl. **11**, 207 (2022).

[25] J. Luo, Z. H. Hang, C. T. Chan, and Y. Lai, Unusual percolation threshold of electromagnetic waves in double‐zero medium embedded with random inclusions, Laser Photon. Rev. **9**, 523 (2015).

[26] J. Bardeen, Electrical conductivity of metals, J. Appl. Phys. **11**, 88 (1940).

[27] F. Sun, J. Fu, J. Sun, Y. Liu, Y. Jin, and S. He, Planar hyper-lens with uniform pre-designed magnification factor by homogeneous medium, Appl. Phys. Express **14**, 022007 (2021).

[28] A. P. Feresidis, G. Goussetis, S. Wang, and J. C. Vardaxoglou, Artificial magnetic conductor surfaces and their application to low-profile high-gain planar antennas, IEEE Trans. Antennas Propag. **53**, 209 (2005).

[29] H. Kamoda, S. Kitazawa, N. Kukutsu, and K. Kobayashi, Loop antenna over artificial magnetic conductor surface and its application to dual-band RF energy harvesting, IEEE Trans. Antennas Propag. **63**, 4408 (2015).

[30] S. B. Glybovski, S. A. Tretyakov, P. A. Belov, Y. S. Kivshar, and C. R. Simovski, Metasurfaces: From microwaves to visible, Phys. Rep. **634**, 1 (2016).

[31] Y. G. Ma, Y. C. Liu, L. Lan, T. T. Wu, W. Jiang, C. K. Ong, and S. L. He, First experimental demonstration of an isotropic electromagnetic cloak with strict conformal mapping, Sci. Rep. **3**, 5, 2182 (2013).

**Competing interests**
The authors declare no competing financial interests.

**Data availability**
The main data and models supporting the findings of this study are available within the paper. Further information is available from the corresponding authors upon reasonable request.

**Acknowledgments**
This work is supported by the National Natural Science Foundation of China (Nos. 12374277, 12274317, 61971300, and 11604292), the Basic Research Project of Shanxi Province (202303021211054), and 2022 University Outstanding Youth Foundation in Taiyuan University of Technology.